\documentclass[10pt]{iopart}
\usepackage{graphicx}

\usepackage{xcolor}

\usepackage{iopams}  
\usepackage{cite}
\begin{document}

\title[Black hole shadow as a \textit{standard ruler} in cosmology]{Black hole shadow as a \textit{standard ruler} in cosmology}

\author{Oleg Yu. Tsupko$^1$, Zuhui Fan$^{2,3}$\\
and Gennady S. Bisnovatyi-Kogan$^{1,4,5}$}

\address{$^1$ Space Research Institute of Russian Academy of Sciences, Profsoyuznaya 84/32, Moscow 117997, Russia}
\address{$^2$ South-Western Institute for Astronomy Research, Yunnan University, Kunming 650500, China}
\address{$^3$ Department of Astronomy, School of Physics, Peking University, Beijing 100871, China}
\address{$^4$ National Research Nuclear University MEPhI (Moscow Engineering Physics Institute), Kashirskoe Shosse 31, Moscow 115409, Russia}
\address{$^5$ Moscow Institute of Physics and Technology, 9 Institutskiy per., Dolgoprudny, Moscow Region, 141701, Russia}

\eads{\mailto{tsupko@iki.rssi.ru}, \mailto{tsupkooleg@gmail.com}, \mailto{zuhuifan@ynu.edu.cn}, \mailto{fanzuhui@pku.edu.cn}, \mailto{gkogan@iki.rssi.ru}}
\vspace{10pt}
\begin{indented}
\item[]September 2019
\end{indented}

\begin{abstract}
Advancements in the black hole shadow observations may allow us not only to investigate physics in the strong gravity regime, but also to use them in cosmological studies. In this paper, we propose to use the shadow of supermassive black holes as a \textit{standard ruler} for cosmological applications assuming the black hole mass can be determined independently. First, observations at low redshift distances can be used to constrain the Hubble constant independently. Secondly, the angular size of shadows of high redshift black holes is increased due to cosmic expansion and may also be reachable with future observations. This would  allow us to probe the cosmic expansion history for the redshift range elusive to other distance measurements. Additionally, shadow can be used to estimate the mass of black holes at high redshift, assuming that cosmology is known.
\end{abstract}

\vspace{2pc}
\noindent{\it Keywords}: black hole shadow, standard ruler, cosmology, Hubble constant, supermassive black holes
%\maketitle

\section{Introduction}

% \textit{Introduction.}

Most galaxies harbour central supermassive black holes. The amazing shadow obtained for the central black hole of M87 has opened up ways to probe strong gravity regimes with high precision \cite{Synge-1966, Bardeen-1973, Luminet-1979, Chandra, Dymnikova-1986, Falcke-2000, Frolov-Zelnikov-2011, Perlick-2014, Perlick-2015, Interstellar-2015, HiokiMaeda2009, Bambi2013, TsukamotoLiBambi2014, Zakharov-2014, Cunha-PRL-2015, Perlick-Tsupko-BK-2015, Rezzolla-Ahmedov-2015, Johannsen-2016, Konoplya-2016a, Perlick-Tsupko-2017, Tsupko-2017, black-hole-cam-2017, Chakraborty-2017, Cite2018-Eiroa, Cite2018-Tsukamoto, Cite2018-Herdeiro, Johannsen-Wang-2016, Lu-Broderick-2014, Lu-Krichbaum-2018, Moscibrodzka-2009, Psaltis-2010, Doeleman-2017, Mizuno-NatAstr-2018,  Giddings-2019, Held-2019, Neves-2019, Gralla-2019, Yunes-2018, Yunes-2019, Dokuchaev-2019, Shipley-2016, Okounkova-2019, Mars-2018}. The shadow of a black hole results from its strong gravitational light bending and the capture of light rays by its event horizon. The current M87 observations have utilized eight world-wide telescopes to form a very long base line interferometry array, reaching an angular resolution of $\sim$25 microarcsec at the wavelength 1.3mm. Going to shorter wavelength, adding new telescopes in the network, or building space-based interferometry in the future can further increase the angular resolution \cite{EHT-1, EHT-2, EHT-3, EHT-4, EHT-5, EHT-6}. This would potentially allow us to observe black hole shadows at cosmological distances, thus making their cosmological applications possible.

The $\Lambda$CDM cosmological model shows a high level of agreement with most of the available observations. Nevertheless, with the increase of observational precision, inconsistencies emerge for some cosmological parameters measured from different observations. For example, there is a $\sim 4.4\sigma$ tension between the local measurement of the Hubble constant employing standardizable candles and that derived from Planck cosmic microwave background radiation observations \cite{cosmology-1, cosmology-2}. Such a tension may ask for physics beyond the $\Lambda$CDM model. To solidify the observed tensions, independent cosmological probes are highly desired. In recent studies by Freedman et al. \cite{Freedman-2019}, they based on a calibration of the Tip of the Red Giant Branch applied to the Type Ia supernovae to determine the distance scale. The derived Hubble constant $H_0=69.8\pm 0.8 (stat) \pm 1.7 (sys) \, \hbox{kms}^{-1}\hbox{Mpc}^{-1}$, in between the Planck result and that from Cepheid calibrations.

% A special role in cosmology is played by objects with certain known, or as usually said, standard properties. Their standard properties allow us to determine the distance to them, which helps us to study and compare different cosmological models. The most famous example of such objects is the so-called standard candles, namely, objects with a known luminosity. Observations of supernovae of Ia type, which are considered as standard candles in cosmology, made it possible to discover the accelerated expansion of the Universe \cite{SN-1998, SN-1999}.

The tremendous advancements in gravitational wave detections and the black hole shadow observation allow us not only to investigate physics in the strong gravity regime, but also to use them in cosmological studies. Many years ago, long before direct detection of gravitational waves, Schutz proposed using sources of gravitational waves as standard sirens in cosmology \cite{sirens-1986}. Subsequently, this was discussed by Holz and Hughes \cite{sirens-2005}. Shortly after the detection of GW170817 from the merger of a binary neutron-star system, measurements of the Hubble constant were carried out using the gravitational wave signals as a standard siren and the redshift from its identified host galaxy taking into account the peculiar motion \cite{sirens-2017}.

In this paper, we propose to use the shadow of supermassive
black holes as a \textit{standard ruler} for cosmological applications. This means that, with a known mass of a black hole, its shadow has a known physical size. With the measured angular size, it can be used to measure the distance. To the best
of our knowledge, it is the first suggestion of use the
shadow in cosmological studies. Considering the
angular size of the shadows, two redshift regimes
can be interesting. At low redshift with $z\le 0.1$, the expected angular size is at the level of $\sim 1$ $\mu$as or larger for black holes with mass $\gtrsim 10^9 M_{\odot}$. \footnote{Note also the recent paper of Mehrgan et al \cite{Mehrgan-2019} where the discovery of supermassive black hole with the mass of $(4.0 \pm 0.80) \times 10^{10} M_{\odot}$ is reported.} Such low redshift distances can be used to constrain the Hubble constant independently. At high redshift with $z>\mbox{a few}$, the shadow size is increased by the cosmic expansion and may also be reachable with future observations \cite{BK-Tsupko-2018}. This would allow us to probe the cosmic expansion history for the redshift range elusive to other distance measurements. \textit{Vice versa}, shadow can be used to estimate the mass of black holes at high redshift, assuming that cosmology is known. Such method could provide better accuracy in comparison with conventional estimates.

\section{Calculation of the shadow size in expanding universe}

To use the shadow of supermassive black hole as a tool for cosmology, we need to understand how the angular size of a shadow depends on its cosmological distance to the observer. In fact, calculating the angular size of a black hole shadow in an expanding universe is not trivial. The main reason is that the trajectories of light rays all the way to the observer are determined simultaneously by the action of the gravity of the black hole and cosmological expansion. This makes it difficult to proceed analytically, and exact analytical solution (valid for arbitrary position of observer, black hole mass and expansion model) is still not found, see discussion in \cite{BK-Tsupko-2018}.

So far, exact analytical solution for the angular size of the shadow is found only in the particular case of black hole in de Sitter model with the cosmic expansion driven by the cosmological constant. To model Schwarzschild black hole embedded in a de Sitter universe, Kottler (also known as the Schwarzschild-de Sitter) spacetime can be used. Angular size of the shadow as seen by static observer was calculated in paper of Stuchl\'{i}k and Hled\'{i}k \cite{Stuchlik-1999}, see also \cite{Stuchlik-2007, Stuchlik-2018} for other types of observers. The first calculation of the shadow angular size as seen by observer comoving with cosmic expansion has been performed by Perlick, Tsupko and Bisnovatyi-Kogan \cite{Perlick-Tsupko-BK-2018}. For influence of $\Lambda$-term on other effects connected with light deflection see, for example, \cite{Rindler-Ishak-2007, Hackmann-2008a, Hackmann-2008b, Lebedev-2013, Lebedev-2016}.

For general case of black hole shadow in expanding Friedmann universe with matter, radiation and Lambda term, one can use the Einstein-Straus model \cite{Einstein-Straus-1945, Einstein-Straus-1946, Schucking-1954, Stuchlik-1984, Balbinot-1988, Schucker-2009, Schucker-2010} and McVittie metric \cite{McVittie1933, Nolan-1, Nolan-2, Nolan-3, Nolan-4}, see also papers \cite{Gibbons-Maeda-2010, Hobson-2012a, Hobson-2012b, Carrera-Giulini-2010, Aghili-2017, Lake-Abdelqader-2011, Piattella-PRD-2016, Piattella-Universe-2016, Faraoni-2017}. Fully analytical calculation of shadow is not trivial in these models (see discussion in \cite{BK-Tsupko-2018}), and the exact analytical solution for the size of the shadow in the general case has not yet been found. Numerical calculation of shadow size using McVittie metric has been performed in \cite{BK-Tsupko-2018}.

There is a well-known effect of increase of apparent angular size of the object if observed by comoving observer in expanding universe \cite{Mattig-1958, Zeldovich-1964, Dashevsk-Zeldovich-1965, Zeldovich-Novikov-book-2}. In modern literature, it is described in terms of so called angular size redshift relation which relates an apparent angular size of the object of given physical size and its redshift, see, for example, \cite{Hobson, Mukhanov-book, Novosyadlyj-book}. Since formula for angular size redshift relation does not take into account the gravity of black hole, it is not possible to use it for exact calculation of shadow with arbitrary position of observer. Indeed, it is assumed that there is a real object with given physical size, and the light rays emitted (or reflected) from the surface of this object propagate through expanding universe all the time, so light propagation is determined by expansion only. However, in the case of black hole shadow formation, light rays propagate through the space-time with a black hole, and the black hole gravity influences their propagation in addition to cosmic expansion. Near the black hole, light rays are strongly affected by its gravity, and a tiny change of parameters can drastically change the photon trajectory, from flyby travelling to capture.

Angular size redshift relation can help to calculate the shadow approximately \cite{BK-Tsupko-2018}.
In realistic situations, we have the following two conditions: the observer is very far from the black hole, at distances much larger than its horizon; the expansion is slow enough and is significant only at very large scales. Under these conditions, we can neglect the influence of the expansion on light ray motion near the black hole, and the effect of the black hole gravity during a long light travel to the distant observer. Therefore we can calculate the shadow in the following way: we first find an 'effective' linear size of the shadow near the black hole and then substitute it into angular size redshift relation. It has been shown in the paper of Bisnovatyi-Kogan and Tsupko \cite{BK-Tsupko-2018} that this approach serves as a good approximate solution for the shadow size in the general case of expanding FRW universe. The validity of the approximation was checked by comparison with exact analytical solution for de Sitter case and with exact numerical calculation in McVittie metric.

Because the (physical) angular diameter distance decreases at redshift $z \gtrsim 1$ in the $\Lambda$CDM model, we expect an increase of the angular size of a shadow at high redshift. At $z\sim 10$, for a black hole with a comparable mass as that of M87, its shadow size is only about one magnitude smaller than that of M87.

\section{Cosmology from the shadow angular size}

The above studies lay a theoretical foundation for using the physical size of a shadow as a standard ruler to measure the cosmological distance. For that, we need independent measurements on the black hole mass. Then with the observed angular size of the shadow, we can derive the angular diameter distance to the black hole.

By definition, the angular diameter distance is:
\begin{equation}
D_A = \frac{L}{\Delta \theta} \, ,
\end{equation}
where $L$ is the proper diameter of the object, and $\Delta \theta$ is the observed angular diameter. This distance is cosmology dependent. In the flat $\Lambda$CDM model, it can be written as
\begin{equation}
D_A(z) = \frac{c}{(1+z) H_0} Int(z) \, ,
\end{equation}
where
\begin{equation}
Int(z) = \int \limits_0^z \left( \Omega_{m0} (1+\tilde{z})^3 + \Omega_{r0} (1+\tilde{z})^4 + \Omega_{\Lambda 0} \right)^{-1/2} d\tilde{z} \, ,
\end{equation}
where $H_0$ is the present value of the Hubble parameter $H(t)$, and $\Omega_{m0}$, $\Omega_{r0}$, $\Omega_{\Lambda 0}$ are the present dimensionless density parameters for matter, radiation and dark energy, respectively.

If for some objects we are able to measure $z$ and $D_A$ independently, we can extract the cosmological parameters from such kind of observations. In \cite{BK-Tsupko-2018} it has been shown that the angular size of a black hole shadow in the expanding universe can be calculated with a high accuracy as\footnote{Well-known formula $\alpha = L/D_A(z)$ for apparent angular size $\alpha$ of object of known physical size $L$ is written for the case when an influence of central gravitating object on light rays propagation is negligible. In case of BH, its gravitation significantly affects the motion of light rays near BH, and formula (\ref{alpha-sh}) provides only approximate solution for angular size of the shadow. It is valid in approximations that observer is far from BH and that the cosmic expansion is negligible near BH. These conditions are justified for all the observed cosmological objects. In such approximations, additional influence of BH gravity on light propagation in expanding universe can be reduced only to use the effective linear size (radius) of the object ($3\sqrt{3}m$) instead of real one ($2m$). Validity of this approximate formula is proven in our previous paper \cite{BK-Tsupko-2018}.}
\begin{equation} \label{alpha-sh}
\alpha_{\mathrm{sh}}(z) = \frac{3\sqrt{3}m}{D_A(z)} .
\end{equation}
Here $\alpha_{\mathrm{sh}}$ is the angular radius of the shadow, $m=GM/c^2$ is mass parameter, with $M$ being the black hole mass.

\begin{figure}
\begin{center}
\includegraphics[width=0.8\textwidth]{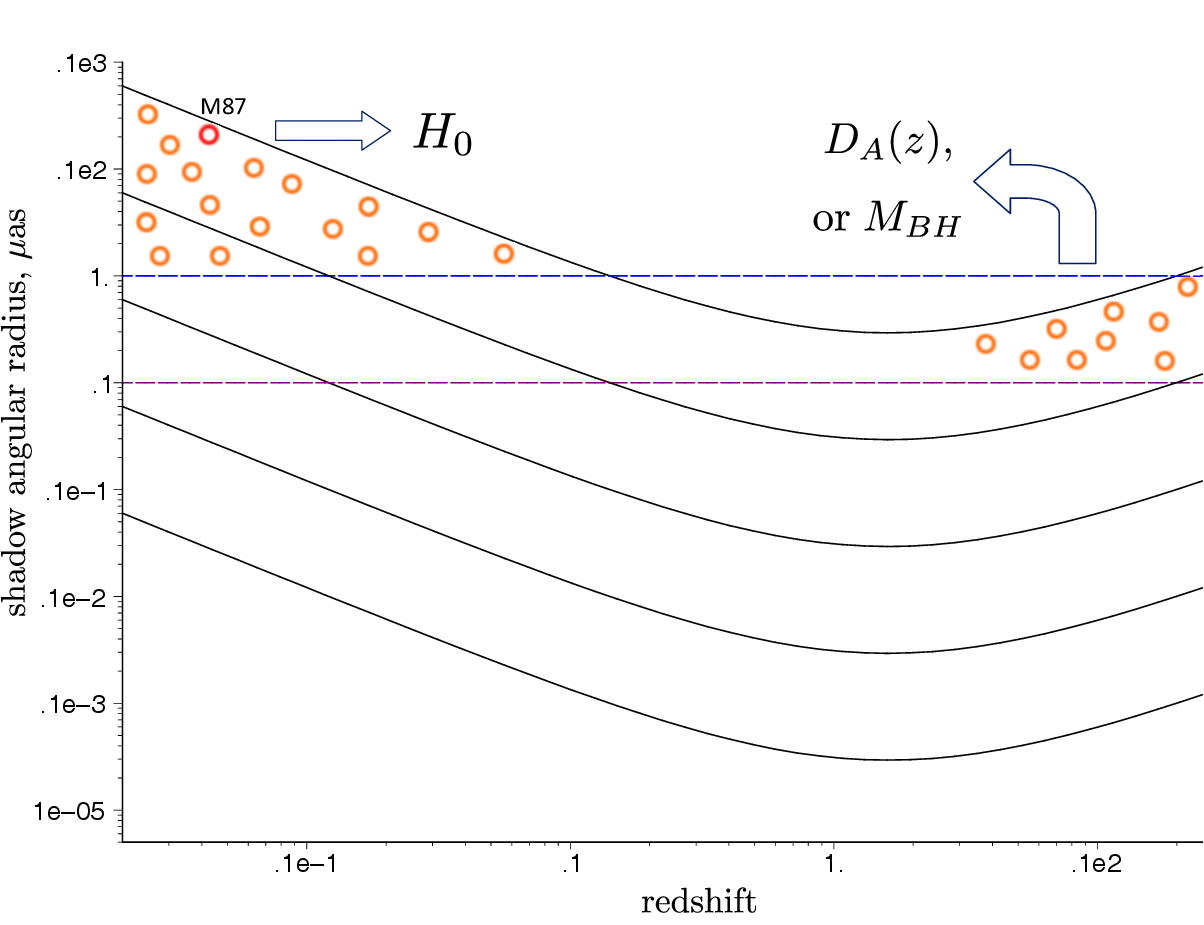}
\end{center}
\caption{(COLOR ONLINE) Using of black hole shadow as a standard ruler in cosmology. Five black curves show predicted angular radius $\alpha_{\mathrm{sh}}$ of black hole shadow for supermassive black holes with masses $10^6$, $10^7$, $10^8$, $10^9$ and $10^{10}$ $M_{\odot}$ (from the bottom to the top) as a function of redshift $z$. They are calculated by formula (\ref{alpha-sh}), for cosmological parameters $H_0 = 70$ (km/sec)/Mpc, $\Omega_{m0}=0.3$, $\Omega_{\Lambda0}=0.7$. Dashed lines show the value $\alpha_{\mathrm{sh}}=1$ $\mu$as (blue) and $\alpha_{\mathrm{sh}}=0.1$ $\mu$as (violet). Orange rings used as a shadow symbol are drawn conditionally, they show in which areas of the parameters to observe to use the method. Observations at small redshifts (left part of figure) with current or one order of magnitude better accuracy could provide a possibility to obtain the Hubble constant. Parameters of black hole in M87, namely $z=0.00428$ \cite{M87-redshift} and $\alpha_{\mathrm{sh}} \simeq 21$ $\mu$as (diameter equals to 42 $\mu$as \cite{EHT-1}), are shown by red ring. Current measurements of the angular size of the shadow along with an independently known distance allowed to obtain the black hole mass in M87 \cite{EHT-1, EHT-6}. Observations at high redshifts (right part of figure) with better sensitivity could provide us a possibility to study cosmology by measurement the function $D_A(z)$, or to estimate the black hole mass, assuming that cosmology is known.}
\label{fig:fig1}
\end{figure}

The effective physical size of a shadow is $3\sqrt{3}m$, which depends only on the black hole mass. If the mass can be determined independently, this physical size can serve as a standard ruler. With the observed angular size $\alpha_{\rm sh}$, we can then measure the angular diameter distance to the black hole. In Fig. 1, we show the $\alpha_{\rm sh}-z$ relation for different black hole masses. The horizontal lines indicate the positions of $\alpha_{\rm sh} = 1 \; \mu\hbox{as}$ and $0.1 \; \mu\hbox{as}$, respectively. From it, we can find two interesting regimes.

1) Nearby galaxies, $z \lesssim 0.1$. For supermassive black holes in nearby galaxies, their shadow radius can be larger than $1 \; \mu\hbox{as}$ if their mass is above $10^9M_{\odot}$. The observed radius for M87 is indicated by the red circle in Fig.1. If we can observe the shadow for a sample of nearby galaxies (illustrated by the orange circles) with independently measured black hole mass, we can then obtain their angular diameter distances to constrain the Hubble constant. 

For small $z$ we have $D_A(z) \simeq cz/H_0$, and we obtain:
\begin{equation}
\alpha_{\mathrm{sh}}(z) = 3\sqrt{3}m \, \frac{H_0}{c} \, \frac{1}{z} \, .
\end{equation}
Therefore the Hubble constant can be found as
\begin{equation}
H_0 = cz \frac{1}{D_A} = cz \frac{\alpha_{\mathrm{sh}}}{3\sqrt{3}m} \, .
\end{equation}

It should be emphasized that to find the Hubble constant, it is not enough to measure the size of only one shadow with a known mass and redshift. The reason is that nearby galaxies experience not only cosmological expansion, but also have their own peculiar velocities, which can make a significant contribution to the redshift. To measure a sample of nearby galaxies at different environments can help mitigate the influence of the peculiar velocities on the $H_0$ estimate. Only in the case that the peculiar motion of a particular galaxy can be well measured, it is possible to remove its effect, and thus to estimate the $H_0$ from a single event (e.g., \cite{sirens-2017}).

2) At large cosmological distances, we can find the angular diameter distance by independently measuring the black hole mass and the angular radius of its shadow:
\begin{equation}
D_A = \frac{3\sqrt{3}m}{\alpha_{\mathrm{sh}}}  \, .
\end{equation}
Peculiar velocities are negligible in comparison with cosmological expansion velocities for high redshifts. If we can also measure the redshift, we can get the dependence of angular diameter distance on redshift, $D_A(z)$, and thus probe the cosmic expansion history at high redshifts.

We note that size of $3\sqrt{3}m$ is derived in the ideal situation when shadow is observed against the background of light sources behind the BH. In the case of accretion flow near BH, the situation is more complicated, and the size of bright ring may be slightly larger, depending on emission profile. In addition, there may be photons from light sources between the observer and the BH that reach the observer and give images of these sources inside the shadow, although they were not affected by the black hole. In this situation, the value $3\sqrt{3}m$ can be considered only as a starting point of investigation. Also, in this paper we have considered only Schwarzschild black hole, whereas the presence of spin may also slightly affect the shadow size, together with deformation of the shadow shape, see, e.g. \cite{Takahashi-2004}. Accurate numerical modelling is required for every individual BH to obtain an exact relation between its mass and the effective linear size of the bright emission ring, as it was done for BH in M87 \cite{EHT-1, EHT-2, EHT-3, EHT-4, EHT-5, EHT-6}.

\section{Observational prospects}

% \textit{Observational prospects.}

The accuracy of using the black hole shadow as a standard ruler is determined by the accuracy of $\alpha_{\rm sh}$ measurements, and the independent black hole mass estimations.

The observed shadow size of M87 has reached an accuracy of about $10\%$ with the angular diameter of $42\pm 3 \; \mu\hbox{as}$ \cite{EHT-1}. As shown in Fig.1, the expected shadow radius is larger than $1 \; \mu\hbox{as}$ up to $z\sim 0.1$ for black holes with mass above $10^{9}M_{\odot}$. To realize such measurements, improvements in the shadow observations are necessary to increase the angular resolution at the level of about one order of magnitude comparing to the current observations. It should be reachable by including multiple space-based telescopes into the ground VLBI array \cite{Fish-2019}. High angular resolution in millimeter wavelengths can be achieved by planned space observatory Millimetron \cite{Millimetron}. In the high redshift regime, an angular resolution of about $0.1 \; \mu\hbox{as}$ is demanded. This can be extremely challenging.  As mentioned in [42], the VLBI technology in optical bands is needed, which can increase the resolution by orders of magnitude due to the shorter wavelengths employed.

For the mass determination of supermassive black holes, there are different ways. The direct methods are based on stellar or gas dynamics near black holes, which have been applied to very nearby ones. For example, see mass estimates for Sgr A$^*$ \cite{Sgr-mass-2008, Sgr-mass-2009} and M87 \cite{M87-mass-2011, M87-mass-2013}. The precision is at the level of about 10 \% without considering the possible large bias between different observations.

Maser observations provide another dynamic approach to determine the central BH mass \cite{Miyoshi-1995, Ho-1998, Kuo-2011, Gao-2016, Darling-2017, Kuo-2018}. The Megamaser Cosmology Project (MCP) \cite{Kuo-2011, Gao-2016} has conducted maser observations for a number of galaxies with redshift up to $z\sim 0.05$, well into the Hubble flow. They can observe not only rotation velocities of masers in the disk near the BH, but also their accelerations by long term monitoring. Thus they can estimate the BH mass independent of the distance to the galaxies. The precision for BH mass estimate can reach a few percent. It is noted that with both the velocity and the acceleration information, the linear size of the disk, and thus the angular diameter distance to the galaxies can be determined. This allows an independent measurement of the Hubble constant \cite{Gao-2016, Reid-2013, Riess-2016}.

We note that to dynamically estimate the BH mass, the distance $D$ to the galaxies is always involved. However, for different observables, the degeneracy between the BH mass and the distance can be different. Considering the Keplerian motion of stars or gas clouds, the velocity measurement leads to $M \propto D$. With the period measurement from astrometric observations, we have $M \propto D^3$. Their combination can give rise to a different degeneracy of $M \propto D^{t}$. In Gillessen et al. \cite{Sgr-mass-2009}, they show $t \simeq 2.19$ for Milky Way BH mass measurement by monitoring the stellar orbits near the BH. In this case, if adding the shadow observations, we have the relation of Eq. (\ref{alpha-sh}) with $M \propto D$. Thus the combination of the shadow data and the stellar motion data can help to break the degeneracy between $M$ and $D$, and the two quantities can be determined simultaneously. For maser observations, although they can determine the mass and distance separately already,  the shadow data can potentially provide an independent observable. The combination of them can expectedly lead to better determination of both the BH mass and the distance $D$.

For higher redshifts, reverberation mapping (RM) method \cite{Blandford-1982, Peterson-2001, Peterson-2019, Shen-2019} can be used, which is another way to determine the mass of black holes dynamically. At the current stage, for RM measurements, the systematic uncertainties of the virial factor can be as large as about 0.3 dex, with its value primarily depending on the bulge properties of AGN host galaxies \cite{Ho-Kim-2014}. Other systematics can also contribute to about a factor of a few to the black hole mass estimate \cite{Shen-2013}. Other indirect methods use correlations between observables and the black hole mass \cite{Peterson-2010, mass-disp-1, mass-disp-2, lum-1, lum-2, lum-3}. Except the very direct dynamic measurements, the uncertainties for the black hole
mass estimate from other methods are still large at the
current stage. With better observations and theoretical understandings, this situation can change and the black hole mass can be better determined. In a recent study by Wang et al. \cite{Jianmin-Wang-2019}, they combine the data from GRAVITY and the reverberation mapping for the AGN broad-line region of 3C 273 to derive simultaneously the BH mass and the distance. Their mass estimate precision is $\sim 50\%$. Such analyses depend on the model of the AGN broad-line region. Adding the shadow data, if achievable, can potentially suppress the influence of systematics involved in the model assumption, thus help the determinations of $M$ and $D$.

At very high redshifts, quasar searches and black hole mass estimates have been done extensively. For example, in paper \cite{Quasar-1}, they reported a quasar at $z=7.085$, and estimated its mass is $M=(2.0^{+1.5}_{-0.7})\times 10^9 M_{\odot}$. In \cite{Quasar-2}, they detected a quasar at $z=7.5$, and the central black hole is estimated to have the mass of $7.8^{+3.3}_{-1.9}\times 10^8 M_{\odot}$. Although the mass estimates have large uncertainties, it has been shown that supermassive black holes with mass of $10^{7-9} M_{\odot}$ already existed at high redshifts. Thus the shadow observations are in principle doable although with great challenges. If the independent black hole mass measurements can be improved, the shadow can be used to probe the cosmic expansion history at very high redshifts. Alternatively, with known cosmology,
the shadow observations provide an independent way to
estimate the black hole mass at high redshifts. This can help us to understand better the physics of supermassive black holes \cite{high-redshift-1, high-redshift-2}.

Admittedly, the method proposed here to use the shadow as a standard ruler is still far from reachable in the near future. Both the shadow observations and the BH mass determinations are very challenging. However, with the fast observational developments, we foresee that the shadow observations can not only provide us knowledge to understand better the BH physics, but also shine light on cosmology in the future.

\section{Concluding remarks}

% \textit{Concluding remarks.}

We have shown that the shadow of a black hole can be used as a standard ruler in cosmology. Two redshift regimes can be interesting considering the redshift dependence of the angular size of the shadow. Shadows of low redshift black holes can be used to constrain the Hubble constant. One order of magnitude better angular resolution would be enough to do that. Shadows of high redshift black holes, with the angular size increased by cosmic expansion, would allow us to probe the cosmic expansion history via independent determination of the angular diameter distance. Additionally, shadows can be used to estimate the mass of black holes at high redshift if the cosmology is known accurately. We emphasize that, unlike known methods where statistical analysis is needed (BAO), our method makes it possible to use an individual object as a standard ruler.

\section*{Acknowledgements}

We are thankful to Prof. Aaron Barth for important comments. Z.F.Fan acknowledges the support from National Natural Science Foundation of China under the grants 11933002, U1931210, 11333001 and 11653001. OYuT would like to thank the SWIFAR visiting fellow program and Prof. Xiaowei Liu and Prof. Xinzhong Er for invitation and kind hospitality during his visit to South-Western Institute for Astronomy Research (SWIFAR) at Yunnan University where this work was started. O.Yu.T. and G.S.B.-K. acknowledges partial support from the Russian Foundation for Basic Research (Grant No. 17-02-00760).

\section*{References}

\bibliographystyle{ieeetr}

\end{document}